# ICT technologies for the refurbishment of wooden structure buildings

Ivan Arakistain, José Miguel Abascal, Oriol Munne

Wood-Biotek Department
Tecnalia Research and Innovation
e-mail: ivan.arakistain@cidemco.es

**Abstract**
Nowadays, one would think that after years of massive concrete and steel construction in Spain, there are not many wood structure buildings left to be refurbished except for some palaces or cathedrals. However, if we go for a walk and have a look at the old part of any city, we will realize that still most of the buildings have a wood structure.

In spite of the fact that the majority of urban regulations forbid their demolition, other bad practices such as casting and overloading the wood structure are very common. Considering that we want to reach a standard of sustainable construction, the economical and environmental costs, which are implied by the deficient refurbishment makes it well worth a previous study of the structure, which in most cases represents less than a 1% of the total budget.

The main goal of this paper is to present most relevant parts of the whole process of diagnosis of a wood structure building by means of Non-Destructive Testing Techniques. Among the ones to be considered, we could mention the analysis of the building and its surroundings, on-site inspection of the building, structural diagnosis, definition of the corrective actions to be taken, definition of treatments, quality control and a maintenance plan.

For the on-site inspection of the building, in the paper we will highlight the use of Non-Destructive methods such as resistograph drilling, X-ray imaging, ultrasound-based testing or moisture measurement. We will provide practical examples of all this.

The aim of this paper is to give the audience an overall idea on how a pre-assessment work can enhance the refurbishment of a wood structure building while reducing costs and environmental impact.

Iván Araquistain, José Miguel Abascal, Oriol Munne

# 1 Introduction

Nowadays, one would think that, after years of massive concrete and steel construction in Spain, there are not many wood structure buildings left to be refurbished except for some palaces or cathedrals. However, if we have a look at the old part of any city, we will realize that still most of the buildings have a wood structure.

Because of their historical value, or preferred location, most urban plans prohibit their demolition, they grant a permission for refurbishment works that, in most cases, are implemented without conducting a thorough study of the building and often result in more expensive works and therefore, we lose part of our historical heritage.

The most common malpractice in wood structure building refurbishment is the emptying of the building, even though the supporting structure of wood is in good condition. We can also highlight the overloading of the supporting structure when trying to adapt it to current regulations or for a change of use.

Assuming that we are within a model of sustainable construction, environmental and economic costs that a poor refurbishment involves well worth conducting a previous study, which in most cases is not even 1% of the project budget.

Therefore, the optimum process of rehabilitation of any wood-structured building requires the fulfilment of a previous assessment work to achieve the following objectives:

-To preserve, the original structure of the building as much as possible by minimizing changes that an adjustment to current regulations and a possible change of use of the building might require.
-Determining the actual state of conservation of the wooden structure of the building.
-Definition of corrective measures compatible for the entire building structure.
-Choosing the most effective treatment method for the wooden structure.
-Definition of a maintenance plan tailored to identified needs.
-Cost savings in implementing the project by optimizing the corrective / remedial actions, protective treatments to apply...

The stages or phases in which we can divide the assessment of the wooden structure of a building are:

- Analysis of the building and its surroundings.
- On-site inspection of the building.
- Diagnosis of the structure.
- Definition of corrective / remedial measures.
- Definition of curative and preventive treatments.
- Implementation of quality control (materials and processes) and developing maintenance plan.

Then, this document will explain the first three steps to diagnose a wooden structure making reference to the most relevant Non-Destructive Techniques which are a complementary tool for the inspector's experience.

Finally, the paper will briefly introduce wireless network-based sensors which will probably gain popularity in the coming years as a tool to predict defects.

# 2 Analysis of the building and its surroundings

## 2.1 Analysis of the building's surroundings

Prior to the analysis of the building itself, there should be an analysis of the surroundings in which relevant aspects are valued for their potential importance in the maintenance of the building:

- Geographical Location.

- Location of building (urban, rural, old town ...).

- Contact with other wood-framed buildings.

- Presence of parks, gardens and forest areas nearby.

- Proximity of rivers and lakes.

- Interior or coastal area.

- Prevailing winds.

These aspects, even though they may not be determining, lead us to search for possible pathologies that we find during the inspection of the building, and we should be very careful about. For example, in a building located in a very rainy environment, we must pay close attention to the status of the deck, looking for possible leaks that may have caused decay in the wood structure. In a building with mountainous areas in the surroundings, we must remain very attentive to the possible presence of termites, reviewing very carefully critical areas (basements, uninhabited and dark areas...).

**2.2 Historical background of the building**

After previous analysis of the building, it is convenient to have a detailed description of the current structural features of the building, including architectural features, building layout and uses to which the building has been allocated until current date.

If there was a rehabilitation project of the building before the study, it is also important to analyze all available information for this new project. Especially when it involves significant structural changes in the building (For example. adding another plant) or when changes arise in the use of the building regulatory requirements involving larger overloads for use, resistance of the structure in fire, soundproofing…

# 3 On-site inspection of the building

The building inspection is aimed at detecting constructive defects in the building that may affect the wood structure, determining the health status of the structure including characterization, evaluation and classification of each of the parts constituting the structure.

The inspection of a wooden structure should be done by qualified and properly trained professionals in the detection, identification and assessment of biotic and abiotic origin damage to wood, as in the identification and measurement of various wood intrinsic singularities (knots, cracks, gems, ...), which are used for structural classification. However, experience requires quite frequently Non-Destructive evaluation techniques.

**3.1 Non-Destructive Testing Equipment**

*-Wood hydrometer:* This equipment is intended to determine on-site, quickly and easily, the wood moisture content. It is based on the electrical conductivity of wood depending on moisture content. See equation (1), where R is the electrical resistance in Ohms and M the moisture content in % [1].

$$Log\ [log(R)\ -4] = 1.009 - 0.0322 M \qquad (1)$$

This is a key factor in the presence of wood xylophagous, especially for fungus, which generally starts to attack wood when moisture content is over 18% - 20%.

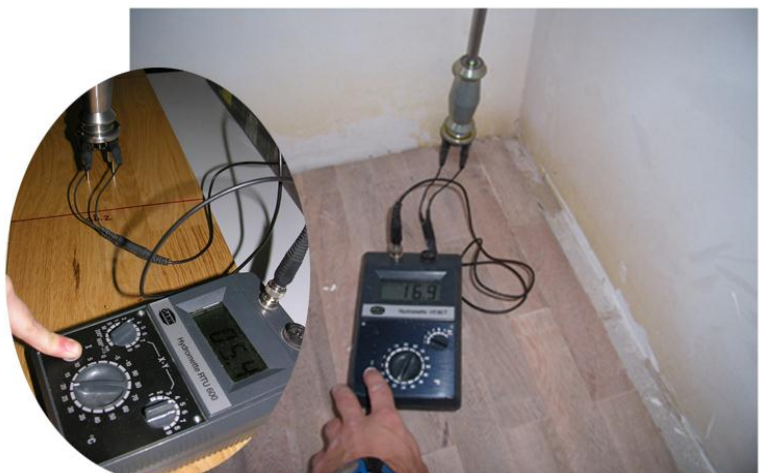

Figure 1: Wood moisture measurement

*Microphotograph:* although it may be relatively simple to determine if a wooden structure was made from one or another sort of wood, it is sometimes very difficult to distinguish all species of wood, especially when they belong to the same family. In such cases, there is no choice, but to take a small sample to identify by means of microphotography.

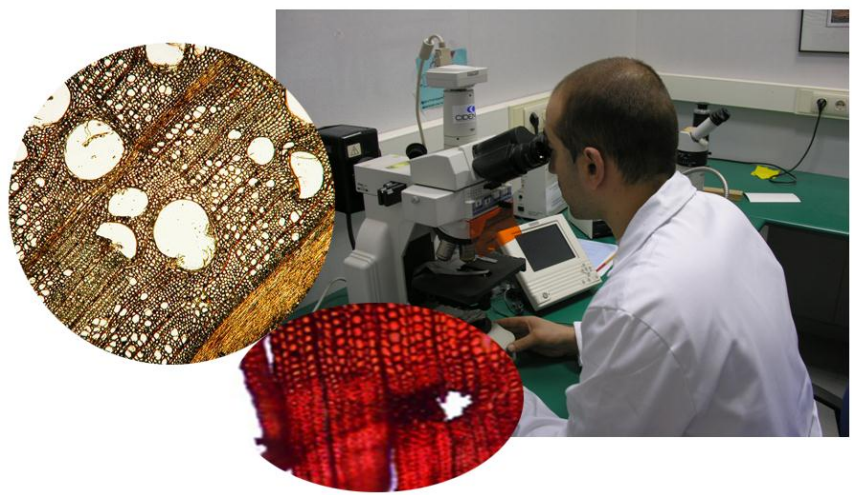

Figure 2: Microphotograph of a wood specie to be identified

*-Resistograph drilling:* This equipment consists of a mechanical drill that evaluates the resistance of the piece of wood to the drilling, allowing detecting losses in wood density, derived from decay or the presence of internal galleries. If we suspect of the existence of hidden cavities or degradations inside a piece of wood (decay when fixed to the wall, termite infested building, resistograph...), we can use the resistograph to verify the integrity of the part.

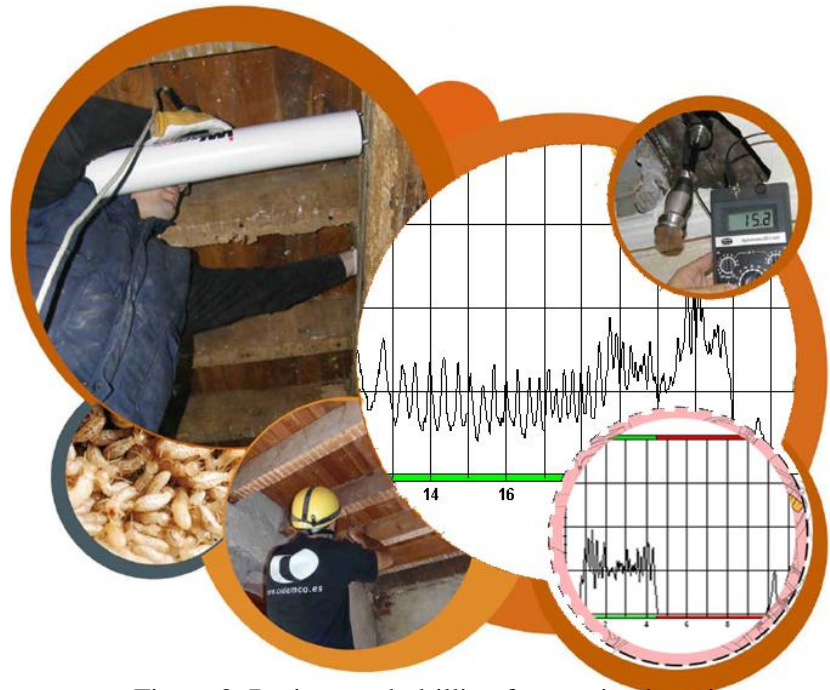

Figure 3: Resistograph drilling for termite detection

Resistograph drilling plays a specially important role, when it is not possible to evaluate the whole structure visually, so as not to provoke any damage in the structure itself.

*-Ultrasonic or induced vibration:* Ultrasonic equipments are used for determining the dynamic modulus of a piece of wood from the speed of wave propagation once the density of wood has been given, but in no case can we determine their resistance. Induced vibrations equipments are based on the same principle, relating the velocity of wave propagation (in this case caused by an impact), with the modulus of elasticity and density of the wood [2].

*-X-ray inspection:* A portable X-ray machine, to determine on site the presence of hidden cavities and the presence of wood-eating organisms inside the wood (Figure. 4).

The earliest literature indicates that the use of the X-rays in the inspection of wood pieces began in the last century, through the qualitative wood evaluation for specific applications, such as tree poles, propellers, and other parts of airplanes in the examination of internal defects [3] (knots, internal cracks, biological decomposition, wood boring insects, etc.). Afterwards, several authors applied X-ray densitometry for the quantitative evaluation of wood properties of different forest species and wood products.

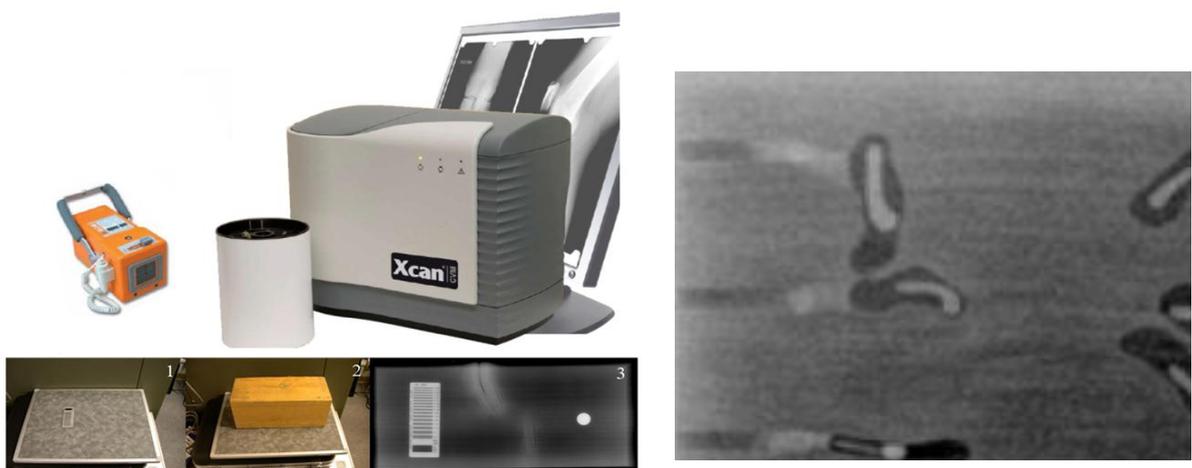

Figure 4: X-ray imaging equipment and image of marine borers in a wood sample

*-GPR:* Ground Penetrating Radar that uses electromagnetic radiation in microwave spectrum allows the prospection of building structures, including wood structures. An antenna located on the material surface radiates energy downwards the material. The response of the material in the form of reflected waves is recorded and analysed so as to evaluate the properties of the medium by which they travel. Some authors have proposed the utility of the GPR to assess wood structures. [4]

*-Infrared Thermography:* Thermography is the use of an infrared imaging and measurement camera to "see" and "measure" thermal energy emitted from an object as it detects IR heat and converts it to electrical signals that are displayed on video screen. These cameras have shown great potential as Non-Destructive evaluation tools. Infrared thermography is becoming a technique that is being used to detect moisture and decay in wood structure buildings. [5]

### 3.2  Determination of health status of the wood structure

The determination of sanitary conditions is to detect, identify and assess damage from biotic and abiotic origin that has affected the structure.

The main abiotic agents that can degrade a wooden structure are:
- Atmospheric agents: solar radiation and humidity changes
- Chemicals: bleach, detergents, acids...
- Mechanical: abrasion, acts of man...
- Fire

Main biotic agents that can cause decay are:

- Mushrooms, xylophages, fungi and decay fungi chromogens
- Marine borers: mollusks and crustaceans
- Insect borers: larval cycle insects (Lepidoptera, Hymenoptera and Coleoptera) and social insects (termites)

The moisture content is a key factor in the presence of organisms attacking wood, especially fungi which usually starts to attack wood above the 18-20% humidity.

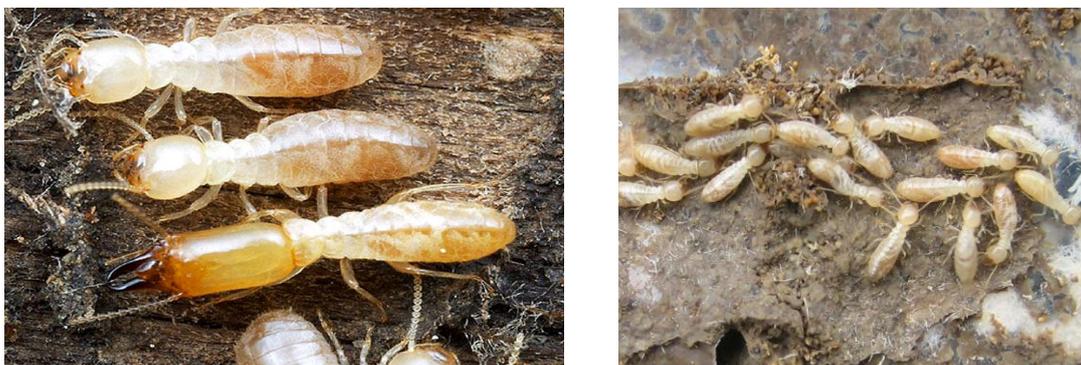

Figure 5: Termites are social insects. In the left image, two worker termites and a warrior can be clearly seen

## 4  Structural diagnosis

To manage an adequate diagnosis of the structure, it is necessary to collect, organize and analyze all the information gathered during the inspection, for which contour maps and tables are developed to facilitate the understanding of the conservation status of the building.

Finally, we must make a structural calculation of the wooden structure based on the completion of the Technical Building Code. The following basic documents are applied:

- Basic Document for Structural Safety in the building DB-SE AE related to actions in the construction.

-Implementation of the Basic Structural Safety Document for the building DB-SE M concerning structural safety in wood.

- Specific models are developed in 2D or 3D for the structural design of each type of building.

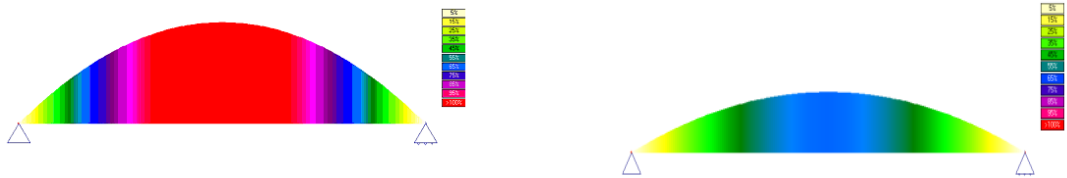

Figure 6: Index of stress of beams, joists and rafters of the wooden structure cover

Load tests are sometimes performed for the diagnosis of wooden structures, which provide far less information than a good structural study and, therefore, are not properly justified.

In case of performing a load test, experts make us aware of the convenience of carrying out a study beforehand to ensure the integrity of the building and safety of the technical staff.

# 5 The future of refurbishment works: Wireless sensor network-based monitoring

### 5.1 The importance of prediction

Many of the structures currently in use were built over 50 years ago. A significant number of the structures are in urgent need of strengthening, rehabilitation or replacement. Many structures are functionally obsolete because they no longer meet current standards. The expensive cycle of maintaining, repairing and rebuilding has led some owners to seek more efficient and affordable solutions to the use of monitoring. Predicting future defects and locating them with wireless sensor networks can save a big sum of money in the maintenance of historical heritage.

### 5.2 Background technology

Ultra low power consuming Wireless Sensor Networks (WSN) are in constant development. These sensors, have a very long battery life (3-5 years or even more) and are intended to monitor several parameters that can be very useful for the maintenance of cultural and historical heritage.
Micro strain, moisture content, air quality, temperature, etc, almost anything can be measured remotely and transmitted to a PC or mobile phone on a very low budget. WSN provide a vital link between monitored structures and a central monitoring site. This allows many structures to be monitored at a central site, with information transmitted via Internet, thereby eliminating costly permanent site installation and reducing the number of site visits.

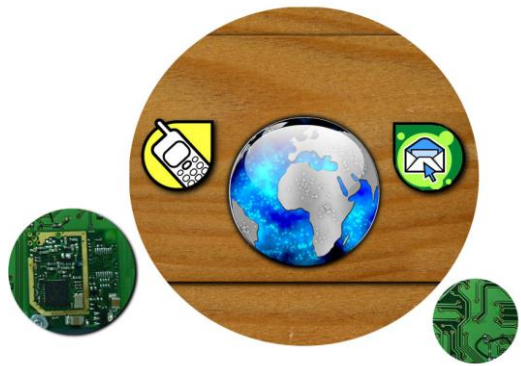

Figure 7: Wireless sensor networks as a tool for the prediction of constructive defects

Apart from monitoring of environmental parameters, wood pieces that set a structure can also be identified by means of new technologies. Radio-frequency identification (RFID) is a technology which replaces bar codes. It is used for automatically identifying an item. To do this, the system relies on RFID tags. These are small transponders that can transmit static information over a short distance, whenever they are asked to. Thanks to RFID tags, each wood piece of a structure can be easily identified and there is not a maintenance problem as they use no battery. [6]

# 6  Conclusion

In brief, assessment work prior to refurbishment is a key activity to maintain our cultural heritage while reducing restoration costs. However, assessment is a challenging activity in need of inputs from Non-Destructive Techniques and wireless monitoring to be used at full capacity. [7]

About 40% of Spanish buildings are over 50 years old, many of them with a wood structure, apart from infrastructure. Restoring them undoubtedly represents a big sum of money, difficult to reach due to the economical crisis. That is why a special effort is needed to improve current inspection procedures, so as to be cost-efficient at restoration works, maintaining as much as possible of our cultural heritage.

Therefore, there is no denying that Non-Destructive Techniques can currently provide relevant information, but not everything is done. The future of refurbishment is undoubtedly linked to ICT development, which will provide new inspection techniques and procedures.